# Efficient statistical analysis of large correlated multivariate datasets: a case study on brain connectivity matrices

Djalel Eddine Meskaldji[1], Leila Cammoun[1], Patric Hagmann[2], Reto Meuli[2],
Jean Philippe Thiran[1] and Stephan Morgenthaler[3]

[1]Signal Processing Laboratory (LTS5), Ecole Polytechnique Fédérale de Lausanne (EPFL), Lausanne, Switzerland.
[2]Departement of Radiology, University Hospital Center and University of Lausanne, Switzerland
[3]Chair of Applied Statistics, Institute of Mathematics, Ecole Polytechnique Fédérale de Lausanne (EPFL), Lausanne, Switzerland.

**Abstract**

In neuroimaging, a large number of correlated tests are routinely performed to detect active voxels in single-subject experiments or to detect regions that differ between individuals belonging to different groups. In order to bound the probability of a false discovery of pair-wise differences, a Bonferroni or other correction for multiplicity is necessary. These corrections greatly reduce the power of the comparisons which means that small signals (differences) remain hidden and therefore have been more or less successful depending on the application. We introduce a method that improves the power of a family of correlated statistical tests by reducing their number in an orderly fashion using our a-priori understanding of the problem . The tests are grouped by blocks that respect the data structure and only one or a few tests per group are performed. For each block we construct an appropriate summary statistic that characterizes a meaningful feature of the block. The comparisons are based on these summary statistics by a block-wise approach. We contrast this method with the one based on the individual measures in terms of power. Finally, we apply the method to compare brain connectivity matrices. Although the method is used in this study on the particular case of imaging, the proposed strategy can be applied to a large variety of problems that involves multiple comparisons when the tests can be grouped according to attributes that depend on the specific problem.

***Keywords and phrases:*** Multiple comparisons ; Family-wise error rate; False discovery rate; Bonferroni procedure; Human brain connectivity; Brain connectivity matrices.

# Introduction

Biomedical research produces large scale datasets which contain multiples markers . This is true for example in genomics with the advent of the gen-chip technology but it is also true in neuroimaging, where functional and structural imaging techniques have made tremendous progress. For all theses techniques their full potential has been held back by what might be called the "curse of multiplicity". In diagnostic applications of medical imaging, it is of interest to compare an image to a standard or to compare two images. If this is done by a pair-wise comparison of voxels, a great number of tests must be performed and if we naively set a test level at $\alpha = 5\%$ for each pair-wise comparison, we would expect 500 false positives per $10'000$ voxels even if no real differences existed. This is called the multiplicity problem and the example shows that it is necessary to control the rate of false positives. The logic of a multiple comparisons situation is summarized in Table 1.

| *Number of null hypotheses which are* | *declared non significant* | *declared significant* | *Total* |
|---|---|---|---|
| *True* | $U$ | $V$ | $m_0$ |
| *False* | $T$ | $S$ | $m_1$ |
| *Total* | $m - R$ | $R$ | $m$ |

Table 1: The general outcome of testing multiple null hypotheses is shown in this table. A total of $m$ null

hypotheses are tested. $V$ is the number of Type I errors or the number of false positives (null hypotheses declared significant). Significance in our sense means active voxels in single-subject experiments or an area that differs between groups. $T$ is the number of Type II errors or the number of false negatives (null hypotheses not declared significant). The power is equal to the probability of detecting a false null hypothesis. The number $R$ of all hypotheses declared significant is an observable random variable, while $S$, $T$, $U$, and $V$ are unobservable random variables. The number of true null hypotheses $m_0$ is also unknown in practice. See Benjamini and Hochberg (1995), Dudoit and van der Laan (2008). The empirical type I error rate is defined by $V/m_0$, while the empirical type II error rate is defined by $T/m_1$.

In neuroimaging most problems involving multiple comparisons use one of two criteria to control the false positives:

- the family-wise error rate (the probability of at least one false positive) or its expected value.
- the false discovery rate (the expected proportion of false positives among all rejections).

Traditional multiple testing procedures focused on controlling the FWER, which can quite easily be achieved via the Bonferroni (1936) correction which consists in dividing the type I error rate of the individuals tests by the number $m$ of tests. The Bonferroni correction has very low power, but exerts a strong control over the false positives which means that the FWER is controlled under any combination of hypotheses ($m_0$ may be different of $m$) unlike weak control when error rates are controlled only under complete null hypotheses ($m_0 = m$). See for example Hochberg and Tamhane (1987). More recently, many corrections for multiplicity were introduced as alternatives to the Bonferroni procedure which were in most cases modifications of the Bonferroni correction such as Sidák (1967) and step-up and step-down procedures, namely, Holm (1979), Simes (1986) and Hochberg (1988). As an alternative to the FWER, Benjamini and Hochberg (1995) introduced the false discovery rate and proposed a procedure to control it based on the procedure of Simes (1986). The FDR has been much used for neuroimaging since it is less restrictive then FWER. Many of these newer methods have been compared in Horn and Dunnett 2004. All these simple alternatives to the Bonferroni correction have effectively critical values which give an increase in terms of power but at the same time, they allow more false positives. This holds in particular for the FDR procedures. Further, all these procedures are ineffective when the number of tests is very big, which is the case in high-resolution neuroimaging. Last but not least, FDR procedures do not have a great advantage over the FWER control procedures with regard to power, when the number of false (null) hypotheses is small compared to the number of all tests which is for example the case in MRI data-sets. See Benjamini and Hochberg (1995), Horn and Dunnett (2004), Logan and Rowe (2004).

One could say that the challenge in multiple testing is to find a method that treats multiple comparisons wisely, which means finding a method that increases the power substantially while at the same time tightly controlling the false positives, similar to Bonferroni . It is necessary to revise the theory of multiple comparisons by taking into account the specifity of the underlying problem.

Effectively, none of the generic multiple testing procedures consider the structural correlations between tests. More elaborate multiple testing procedures incorporating such information have been proposed. Examples are the re-sampling techniques introduced in Westfall and Young (1993) based on the estimation of the distribution of the maximum statistic, or application of the random field theory by Worsley et al. (1996) using the expected Euler characteristic. These methods are in general complicated and do not lead to a great increase of the power in most practical cases. See Logan and Rowe (2004).

In the spirit of the remark by Sarkar and Heller (2008) who wrote 'A more explicit use of the dependence structure should result in a powerful method', we propose in this paper a simple and yet effective procedure. In neuroimaging and in many other areas of application, measures are often correlated spatially and quite well defined anatomically. Imaging voxels belonging to a unique anatomical region will usually exhibit correlated behavior whether the measure is structural or functional. Considering adjacency matrices representing brain connectivity networks, connectivity weights of neighboring connections are likely correlated because of their anatomical proximity. We investigate in this paper a simple approach that takes into account this correlation based on a construction of an appropriate statistic in each anatomical or functional block defined by specific attributes of the problem at hand. This statistic summarizes all the information in this block and will be used for comparisons instead of the values observed in this block. Thus, the number of tests will be significantly reduced. At the same time the noise variance will be reduced, which in turn increases the power. Once a block is detected as significant, a local investigation can be performed to precisely locate significance. We show that this method has great potential, in particular in cases where the blocks are well defined and the summary statistics are properly chosen.

# General formulation and notation

Consider a set $R = \{r_j | j = 1, \dots, M\}$ of $M$ pixels/voxels or any surface/volume unit of a 2D/3D image. We call these elements *small regions*. The set $R$ is *the global region of interest* where the comparisons are to be done or which is to be compared between individuals. In each small region $r_j,\ j = 1, \dots, M$, we observe a vector $\boldsymbol{x_j}$ (of dimension $q$) of measures that may be obtained from the image or by direct methods. The original data in the global region of interest is of dimension $M \times q$. We group the $M$ small regions into $m$ blocks $B_1, \dots, B_m$ such that $\bigcup_{i=1}^{m} B_i = R$ and $B_i \cap B_k = \emptyset$ for $i \neq k$, where a block contains a set of small regions that are linked by attributes associated with the problem at hand. The number of small regions in the block $B_i$ is $b_i = |I_{B_i}|$ where $I_{B_i} = \{j | r_j \in B_i\}$, that is

$$\sum_{i=1}^{m} b_i = \sum_{i=1}^{m} |I_{B_i}| = M.$$

Here, |●| denotes the number of elements of the set ●.

*Example 1*
*The global region of interest R is the part of a 3D image that represents the human brain obtained for examples by Magnetic Resonance Imaging (MRI), Diffusion Tensor Imaging (DTI) or Diffusion Spectrum Imaging (DSI). The small regions are in this context the volume units or voxels. To obtain the global region of interest we apply a mask to the entire image such that only the small regions inside the brain are considered. This will improve the power of testing since the number of tests is reduced. See Dudoit and van der Laan (2008). In each small region several measures can be taken (e.g. functional MRI measuremants, fractional anisotropy (FA), diffusion tensor orientation, etc.). The small regions are grouped into blocks defined by a segmentation atlas of the anatomical brain regions.*

For each block $B_i$ we consider an univariate summary statistic $t_i$ (or a p-variate summary statistic $\boldsymbol{t_i}$), a function of all observed data within the block, that is

$$t_i = t_i(\{\boldsymbol{x_j} | j \in I_{B_i}\}), \qquad \text{for all } i = 1, \dots, m,$$

or (in the multivariate case)

$$\boldsymbol{t_i} = \begin{pmatrix} t_i^1 \\ \vdots \\ t_i^p \end{pmatrix} = \begin{pmatrix} t_i^1(\{\boldsymbol{x_j} | j \in I_{B_i}\}) \\ \vdots \\ t_i^p(\{\boldsymbol{x_j} | j \in I_{B_i}\}) \end{pmatrix}, \qquad \text{for all } i = 1, \dots, m.$$

*Remark 1*
*This procedure is akin to applying a smoothing in these blocks but in contrast to the usual smoothing methods, the sizes of the smoothing windows are the sizes of the blocks and thus depend on the nature of the problem and we also never smooth across boundaries.*

The advantages of this method are twofold.

- By pre-selecting the blocks to perform the reduction in complexity, we explicitly take into consideration the structural correlation between the small regions in each block. This is an advantage of the block-wise method. And, of course, the number of block-wise tests $m$ is usually much reduced compared to the number of pairwise tests $M$ of small regions. This is the first reason why the power of testing will be improved.
- It is true that the individual values in the small regions are lost using this procedure, but in the applications we have in mind, detecting significant results in small regions is not of interest. Smoothing will reduce the variance of the outcome, which facilitates the detection of significant structures, while isolated significant small regions are rarely considered. This leads to a desirable robustness of the method and is the second reason for increased power.

*Example 2*
*The global region is a matrix of size* $M = 8 \times 8 = 64$ *which represents observed values of a z-test when we compare neuroimages pixel-wise. All the entries of the matrix are simulated realizations of independent normal random variables with variance 1. This matrix is partitioned into* $\mathrm{m} = 6$ *blocks as indicated in Table 2 where only the top-left block represents a true effect with mean-shift* $\Delta = 3$ *and variance* $\sigma^2 = 1$. *In all other small regions, the mean is 0 except for some entries where the values are simulated outliers.*

| | *1* | *2* | *3* | *4* | *5* | *6* | *7* | *8* |
|---|---|---|---|---|---|---|---|---|
| *1* | ***3.26*** | ***4.48*** | *2.27* | *-0.83* | *0.06* | *0.32* | ***6.32*** | *1.07* |
| *2* | *2.17* | ***4.26*** | ***4.21*** | *0.67* | *-1.39* | *0.69* | *-0.62* | *-1.1* |
| *3* | ***4.48*** | *1.47* | *2.1* | *-2.58* | *1.36* | ***6.23*** | *0.72* | *0.46* |
| *4* | *2.89* | *2.74* | *1.74* | *0.86* | *2.2* | *1.01* | *0.5* | *-1.79* |
| *5* | *-0.29* | *1.06* | *2.73* | *-0.49* | *1.13* | *0.72* | ***9.18*** | *-1.73* |
| *6* | *0.22* | *-0.28* | *-0.16* | *0.45* | *-5.45* | *-0.7* | *-0.19* | *-1.27* |
| *7* | *-0.49* | *0.51* | *-0.64* | *0.2* | *0.44* | *0.18* | *-0.63* | *0.59* |
| *8* | *-1.87* | *1.29* | *-0.23* | *0.6* | *1.37* | *1.94* | *-1.91* | *-0.33* |

Table 2: Partition of the 8 × 8 area into 6 blocks. The entries are simulated as described in Example 2. The significant values after Bonferroni correction are in bold.

*For each entry of the global matrix, we want to perform a small region-wise (SRW) comparisons, that is, to test* $H_j^0: \mu_j = 0 \; vs. \; H_j^1: \mu_j > 0, \; (j = 1, \ldots, 64)$ *at global level* $\alpha = 0.05$. *If the hypotheses are tested simultaneously and the Bonferroni correction is used, the critical value of each test is*

$$c_j = \Phi^{-1}(1 - \alpha/M) = \Phi^{-1}(1 - 0.05/64) = 3.16,$$

*where* $\Phi$ *is the cumulative probability function of the standard normal distribution. In this case only values higher than 3.16 (which are in* ***bold*** *in Table 2) are detected as significant small regions. The empirical type I error rate is* $3/52 = 5.76\%$, while the empirical type II error rate is $7/12 = 58.3\%$.

*Next, we apply the strategy based on blocks. In each block* $B_i$ *we construct a summary statistic based on these different statistics:*

- *the mean of the values observed in the block* $B_i$,
- *the median of the values observed in the block* $B_i$,
- *the Huber (1964) robust estimator of the values observed in the block* $B_i$.

*The results of these summary statistics are presented in Table 3.*

| | | | |
|---|---|---|---|
| *Mean* | ***3.08*** | *0.04* | ***1.15*** |
| | *0.15* | *-0.02* | *-0.12* |
| *Median* | ***2.99*** | *0.36* | *-0.59* |
| | *-0.2* | *0.32* | *-0.26* |
| *Huber mean* | ***3.08*** | *0.07* | *0.54* |
| | *0.07* | *-0.02* | *-0.12* |

Table 3: The application of the block strategy based on three different summary statistics. For each summary statistic we obtain a matrix of size $m = 2 \times 3 = 6$.

*Individual tests in each block* $B_i$ *correspond to test* $H_i^0\colon \mu_i = 0\ vs.\ H_i^1\colon \mu_i > 0,\ (i = 1, \dots, 6)$ *and the critical value of each test if the mean is used as a summary statistic, is*

$$c_i = \frac{1}{\sqrt{b_i}} \Phi^{-1}\left(1 - \frac{0.05}{6}\right),$$

*where* $b_i$*, the size of the block* $B_i$*, is either 8 or 12. The additional factor* $\frac{1}{\sqrt{b_i}}$ *is due to the reduction of the variance by averaging in each block. Although the other two estimators have a variance greater than the variance of the mean, we use the same critical values as those corresponding to the mean for simplification which leads to a conservative procedure. The critical values are 0.691 for the blocks of size 12 and 0.846 for the blocks of size 8. The significant blocks are in* ***bold*** *in Table 3. The empirical type I error is 1/5 for the mean and equal to* 0 *for the robust statistics (median and Huber mean). The empirical type II error is in all cases equal to 0.*

# Properties of block-wise analysis using the block mean as a summary statistic

In this section we investigate the properties of a particular case of the method introduced in the last section. The summary statistic we consider in this section consists in calculating the mean of the values observed in all the small regions within each block. We call this procedure the mean block-wise analysis (mean-BWA) from now on.

We compare this methodology to the approach that separately tests small regions (SRW) in terms of power and type I error rates. Mean-BWA and SRW solve different problems. While mean-BWA tests the significance of blocks, SRW tests the significance of small regions. This has to be kept in mind when comparing level and power of these two methods.

We first consider the ideal case where all the small regions inside a block are affected and then generalize to a the case where only a fraction of the small regions in an affected block are themselves affected.

## *In affected blocks, all small regions are affected*

Denote by $x_j$ for all $j = 1, \dots, M$ the measurements associated with the small regions. In the *non affected small regions* (where the null hypothesis holds) the observed values are of the form $x_j = \mu_0 + \sigma_0 e_j$ and in the *affected small regions* (where the alternative hypothesis holds) the observations are of the form $x_j = \mu_0 + \Delta + \sigma_1 e_j$, where $e_1, \dots, e_M$ are independent realizations of a normal random noise variable with continuous cumulative distribution $\Phi$ (with mean 0 and variance 1). Thus, $\Delta$ is the raw effect. The number of non affected small regions (number of true null hypotheses) within the global region of interest is $M_0$ and the number of affected small regions (number of false null hypotheses) within the global region of interest is $M_1 = M - M_0$. We assume that the variances ${\sigma_0}^2$ and ${\sigma_1}^2$ are known. When proceeding according to SRW we perform a one sided test $H_j^0 : \Delta = 0\ vs.\ H_j^1 : \Delta > 0$, for all $j = 1, \dots, M$ at global level $\alpha$ (i.e. $FWER \le \alpha$ or $FDR \le \alpha$).

Under the Bonferroni correction each single test is performed at level $\alpha/M$. The null hypothesis $H_j^0$ is thus rejected if the observation $x_j > c_j$, and $c_j$ is such that

$$P_{H_j^0}(x_j > c_j) = 1 - P_{H_j^0}\left(\underbrace{\left(\frac{x_j - \mu_0}{\sigma_0}\right)}_{e_j} \leq \left(\frac{c_j - \mu_0}{\sigma_0}\right)\right) = \alpha_j = \frac{\alpha}{M}$$

$$\Rightarrow \frac{\alpha}{M} = 1 - \Phi\left(\left(\frac{c_j - \mu_0}{\sigma_0}\right)\right).$$

This implies

$$c_j = \mu_0 + \sigma_0 \Phi^{-1}\left(1 - \frac{\alpha}{M}\right).$$

When applying themean-BWA, the small regions are grouped into blocks accoring to their physiological or functional attributes so that the number of completely affected blocks obtained is $m$. In each block $B_i$, of size $b_i$ we construct a summary statistic $t_i$ based on the block mean, that is,

$$t_j = \left(\frac{1}{b_i}\right)\left(\sum_{\{j \in I_{B_i}\}} x_j\right) = \bar{x}_i, \quad (i = 1, \ldots, m).$$

Because there are no outliers in this case, the mean will give more or less the same results as one would obtain with a robust estimator.

In the *non affected blocks* which contain only non affected small regions, the statistic $t_i$ is a realization of a random variable $T_i = \mu_0 + \sigma_0 \bar{e}_i$, while in the *affected blocks* $t_i$ is a realization of a random variable $T_i = \mu_0 + \Delta + \sigma_1 \bar{e}_i$ where $\bar{e}_i = \left(\frac{1}{b_i}\right)\left(\sum_{\{j \in I_{B_i}\}} e_j\right)$. The number of non affected blocks is $m_0$ and the number of completely affected blocks is $m_1 = m - m_0$. The distribution of $T_i$ depends on the distribution of $\bar{e}_i$. In the normal case, the distribution of $\bar{e}_i$ is also a normal distribution. In the general case, if the sizes $b_i$ or the number of subjects to be compared are large enough, the central limit theorem (CLT) leads to an approximation of the distribution of $\bar{e}_i$ by a normal $N\left(0, \frac{1}{b_i}\right)$ distribution. We assume from now that the normal approximation may be used.

Fixing the FWER at $\alpha$ and again using the Bonferroni correction, the null hypothesis is rejected in the block $B_i$ if $t_i > c_i$, and $c_i$ is such that

$$P_{H_0^i}(t_i > c_i) = 1 - P_{H_0^i}(t_i \leq c_i) = \alpha_i = \frac{\alpha}{m}$$

$$\Rightarrow \frac{\alpha}{m} = 1 - P_{H_0^i}\left[\left(\underbrace{\frac{\sqrt{b_i}}{\sigma_0}(t_i - \mu_0)}_{\bar{e}_i}\right) \leq \left(\frac{\sqrt{b_i}}{\sigma_0}(c_i - \mu_0)\right)\right]$$

$$\Rightarrow \frac{\alpha}{m} \simeq 1 - \Phi\left(\frac{\sqrt{b_i}}{\sigma_0}(c_i - \mu_0)\right).$$

So, for fixed α,

$$c_i = \mu_0 + \frac{\sigma_0}{\sqrt{b_i}} \Phi^{-1}\left(1 - \frac{\alpha}{m}\right).$$

This relation shows that the critical values for the mean-BWA not only depend on the ratio $(M/m)$ which represents the reduction of the number of tests, but also on $\left(\frac{1}{\sqrt{b_i}}\right)$ which decreases as long as the block size increases and is due to the reduction in the noise variance. This means that for a fixed block segmentation (fixed number of blocks), the power increases with the resolution of the image as long as the signal to noise ratio does not increase at the same time.

We can also compare the powers of the two tests. In the case of SRW the power $1-\beta_j$ can be evaluated as follows

$$
\begin{aligned}
1-\beta_j &= P_{H_1^j}\left(x_j > c_j\right) \\
&= 1 - P_{H_1^j}\left(\left(x_j - (\mu_0 + \varDelta)\right)/\sigma_1 < \left(c_j - (\mu_0 + \varDelta)\right)/\sigma_1\right) \\
&= 1 - \Phi\left(\left(c_j - (\mu_0 + \varDelta)\right)/\sigma_1\right) \\
&= 1 - \Phi\left(\left(\mu_0 + \sigma_0 \Phi^{-1}\left(1 - \frac{\alpha}{M}\right) - \mu_0 - \varDelta\right)/\sigma_1\right) \\
&= 1 - \Phi\left(\left(\sigma_0 \Phi^{-1}\left(1 - \frac{\alpha}{M}\right) - \varDelta\right)/\sigma_1\right).
\end{aligned}
$$

Likewise, for mean-BWA we find

$$1-\beta_i \simeq 1 - \Phi\left[\left(\sigma_0 \Phi^{-1}\left(1 - \frac{\alpha}{m}\right) - \varDelta\sqrt{b_i}\right)/\sigma_1\right].$$

We have $1 - \left(\frac{\alpha}{m}\right) < 1 - \left(\frac{\alpha}{M}\right)$ $and\ \varDelta < \varDelta\sqrt{b_i} \Rightarrow \left(\sigma_0 \Phi^{-1}\left(1 - \frac{\alpha}{m}\right) - \varDelta\sqrt{b_i}\right)/\sigma_1 < \left(\sigma_0 \Phi^{-1}\left(1 -\right.\right.$ $\alpha M - \varDelta/\sigma 1.$ Then $1-\beta i > 1-\beta j$ since $\Phi$ is an increasing function.

Of course, this behavior is partly due to the fact that we are in the ideal situation where affected blocks consist only of affected small regions. In a more realistic situation, we cannot expect the blocks to match the affected regions perfectly. In addition, the power has a different meaning in the two cases. For SRW we want to detect signals on the small region level, while for mean-BWA we seek significant blocks.

We simulated the tests when the observed values in non affected small regions are independent realizations of $x_j \sim N(0,1)$. The observed values of small regions within affected blocks are independent realizations of $x_j \sim N(\Delta, 1)$. For simplicity, the $m$ blocks are chosen to have the same size $b$.

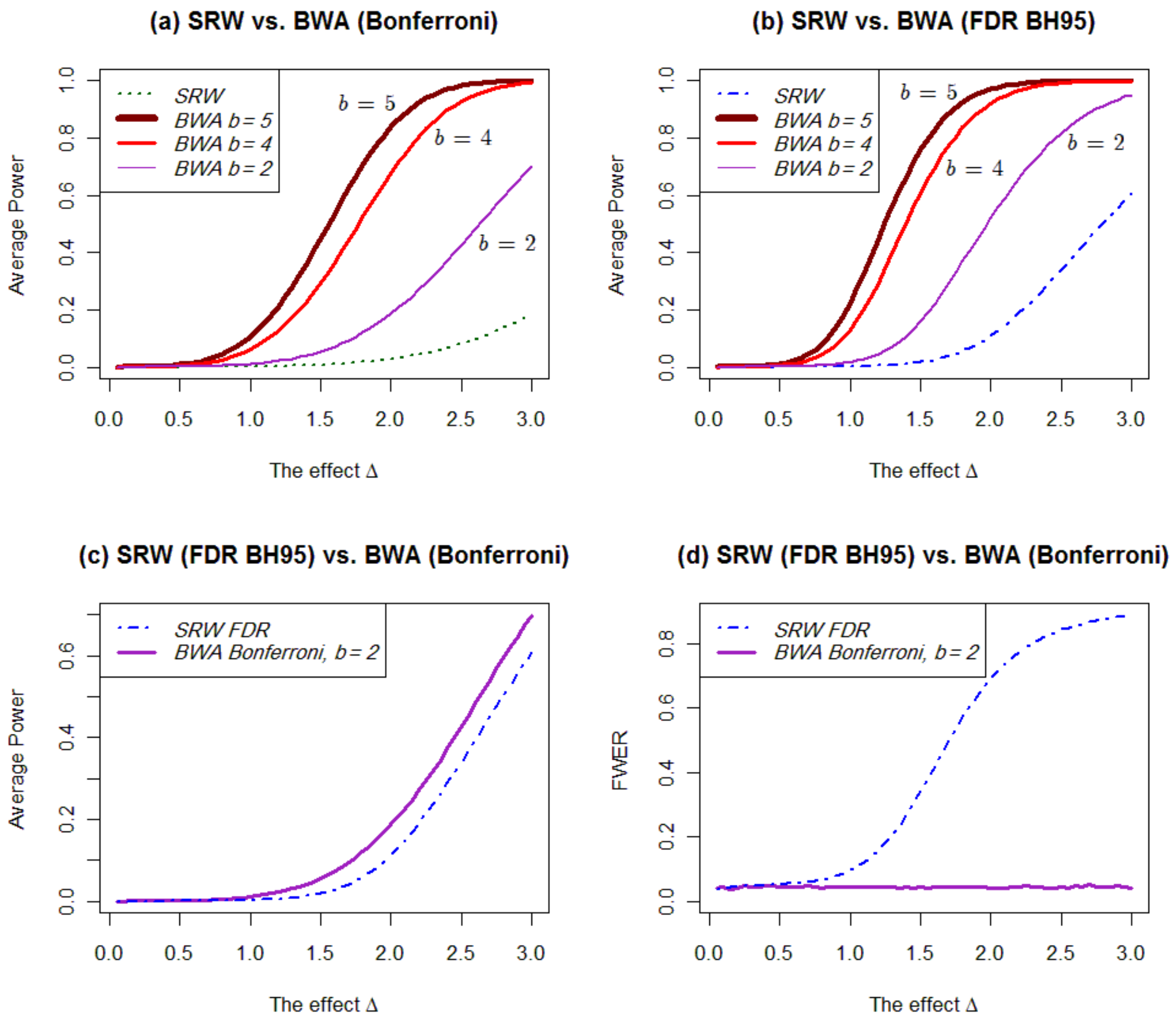

Figure 1: Power and FWER of different methods as a function of the raw effect $\Delta$. (a) Power of mean-BWA ($b = 5$ (dark red), $b = 4$ (red), $b = 2$ (purple)) and SRW (small green points) with the Bonferroni correction for both. (b) Power of BWA ($b = 5$ (dark red), $b = 4$ (red), $b = 2$ (purple)) and SRW (dashed blue) with FDR (BH95) for both. (c) Power of mean-BWA ($b = 2$ (purple)) with the Bonferroni correction and SRW (dashed blue) with the FDR BH95 procedure. (d) FWER of mean-BWA ($b = 2$ (purple)) with the Bonferroni correction and SRW (dashed blue) with the FDR BH95. The other parameters are $M = 1000,\ M_1 = 100,\ \alpha = 0.05$.

Figure 1 shows the behavior of the power of detecting a significant effect. Figure 1 (a) shows the power of detecting significant blocks compared to the power of detecting affected small regions using the Bonferroni procedure. Figure 1 (b) is identical to Figure 1 (a) except that the Benjamini & Hochberg procedure is used (BH95). In Figure 1 (c), we use the BH95 for SRW and the Bonferroni procedure for the mean-BWA with blocks of size 2. The power is estimated as the average power (Kwong et al., 2002), that is, the average of $S/M_1$ (see Table 1) across the simulations. As $\Delta_i = \Delta$ in all affected small regions and the blocks have the same size, the average power is equal to the per-pair power defined by Einot and Gabriel (1975). For the block-wise tests, we obviously have to replace $M$ by $m$ and $M_0$ by $m_0$. Figure 1 (d) shows the estimated FWER of the two strategies used in Figure 1 (c). These estimations are obtained by averaging over $10'000$ simulations.

Generally, the power of detecting a significant effect increases with $\Delta$. This holds for all the methods. The BH95 procedure does not have a big advantage over the Bonferroni procedure for small raw effects $\Delta$. But what is more interesting is the fact that for a fixed significant effect the power of the mean-BWA method increases with the size of the block. When the block size $b = 5$ the power is rapidly increasing towards 1 with $\Delta$. Even when the block size $b$ is equal to 2 which is the worst case for mean-BWA, it still has an advantage over SRW controlled by BH95. We also see in plot (d) of Figure 1 that the FWER of the mean-BWA based on the Bonferroni correction is controlled at level $\alpha\left(\frac{m_0 b}{mb}\right) = \alpha\left(\frac{m_0}{m}\right) = \alpha\left(\frac{M_0}{M}\right)$ which is the expected FWER of SRW controlled by the Bonferroni procedure. This is the desired property that we have discussed in the introduction. The FDR procedures on the other hand allow more false positives than the FWER procedures. With $\Delta > 2$, the estimated FWER of BH95 is in fact close to 1, which means that almost surely at least one declared significance is false and we do not know which one. See Manly et al. (2004) for similar observations. Since the control of FDR is not the same as FWER, this should not be surprising. See Nichols and Hayasaka (2003).

### *In affected blocks, only a fraction of the small regions is affected*

The situation we have discussed so far represents an ideal case for the mean-BWA. It would in fact be more realistic to consider *partially affected blocks*. This could happen for example if the block's limits do not exactly match the true limits of the anatomical or functional blocks (quality of segmentation) or simply, this could happen if only a proportion of small regions in a block were affected. In this section, we consider the effect of this change on the power of detecting a positive effect in a block. Consider a partially affected block of size $b_i$ where only $k_i$ ($k_i = 1, \dots, b_i$) of the small regions are affected. In a such block, there are $k_i$ observed values of the form $x_j \sim N(\mu_0 + \Delta, \sigma_1^2)$ and $(b_i - k_i)$ are of the form $x_j \sim N(\mu_0, \sigma_0^2)$. So, in each partially affected block $B_i$,

$$x_j \sim N\left(\mu_0 + \frac{k_i}{b_i}\Delta, \sigma_{k_i}^2\right)$$

where

$$\sigma_{k_i}^2 = \frac{k_i}{b_i}\sigma_1^2 + \left(1 - \frac{k_i}{b_i}\right)\sigma_0^2 + \Delta^2 \frac{k_i}{b_i}\left(1 - \frac{k_i}{b_i}\right).$$

In this case, the power of detecting a significant block effect using mean-BWA is

$$1 - \beta_i = 1 - \Phi\left(\left(\sigma_0 \Phi^{-1}\left(1 - \frac{\alpha}{m}\right) - \frac{k_i}{b_i}\Delta\sqrt{b_i}\right)/\sigma_{k_i}\right),$$

As $\frac{k_i}{b_i}\Delta\sqrt{b_i} < \Delta\sqrt{b_i}$, the power of the mean-BWA procedure is reduced.

For the simulations we chose $\mu_0 = 0,\ \sigma_0 = \sigma_1 = 1,\ b_i = b$ for all $i$ in $\{1, \dots, m\}$ and $k_i = k$ for all partially affected blocks. If the number of completely affected blocks is $m_1$ and the number of partially affected blocks is $m_2$, the total number of affected small regions is $M_1 = (m_1 + \frac{k}{b}m_2)b$. In this case we have ${\sigma_k}^2 = 1 + \Delta^2 \frac{k}{b}\left(1 - \frac{k}{b}\right)$ and the power of detecting a significant effect in a partially affected block is given by

$$1 - \beta_i = 1 - \Phi\left(\left(\Phi^{-1}\left(1 - \frac{\alpha}{m}\right) - \frac{k}{b}\Delta\sqrt{b}\right)\Big/\sqrt{1 + \Delta^2 \frac{k}{b}\left(1 - \frac{k}{b}\right)}\right).$$

In Figure 2 we can see the behavior of the power of the three different multiple testing procedures (SRW with Bonferroni, SRW with BH95 and mean-BWA with Bonferroni) depending on the proportion of truly affected small regions $k/b$ for two different values of the significant effect $\Delta$ and two different values of the block size $b$. The power is obtained by simulations of random normal samples. In this case, the power of mean-BWA procedure is evaluated as the power of detecting a significant effect in an entire block where only a proportion $k/b$ of small regions are truly affected.

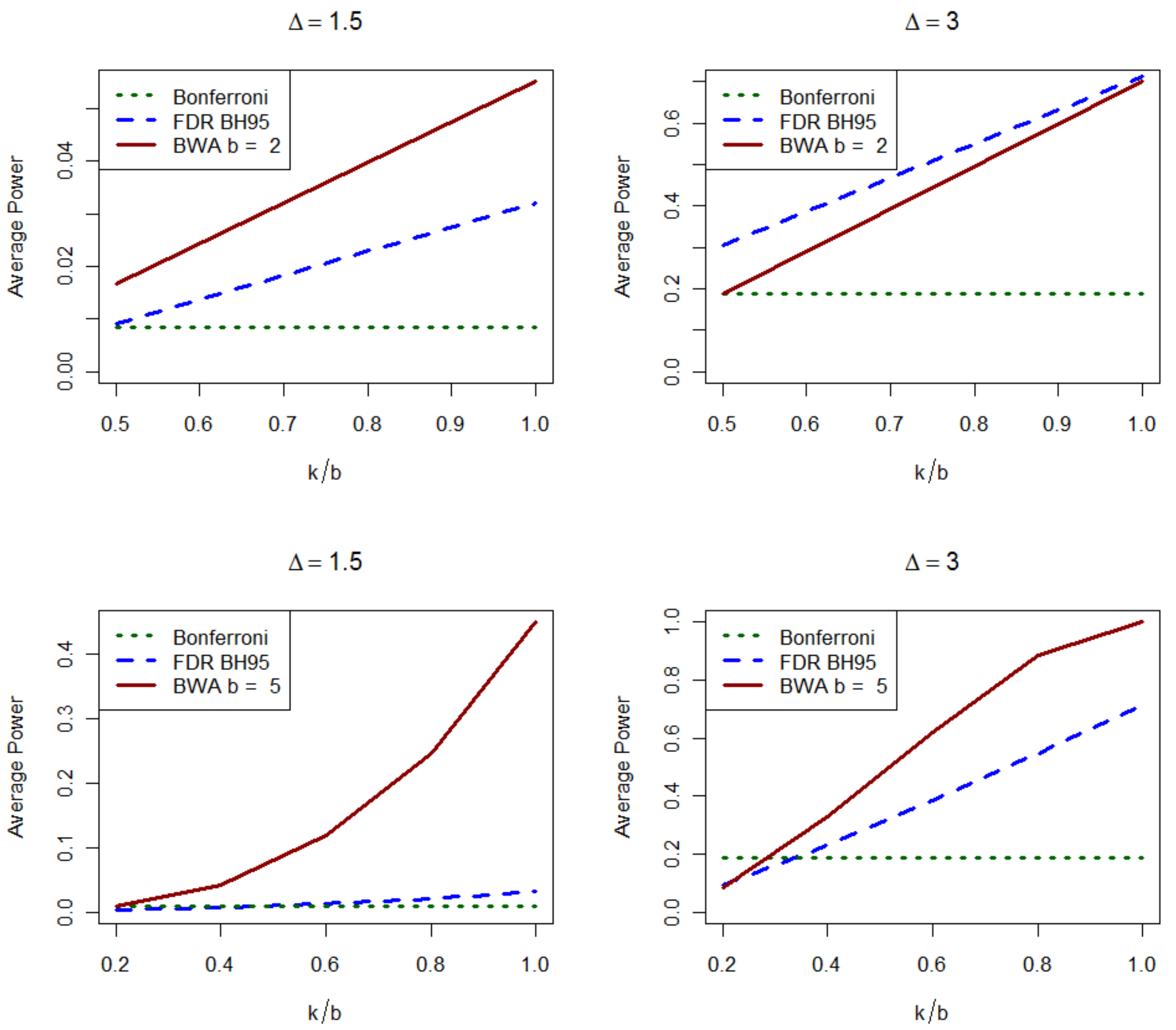

Figure 2: Power of mean-BWA with Bonferroni correction (continuous line) compared to SRW with BH95 correction (dashed blue line) and Bonferroni correction (green points) for different values of $\Delta$ and $b$ depending on the proportion $k/b$ of affected small regions within each partially affected block. The other parameters are: $M = 1000,\ m_1 = 0,\ m_2 = m/5,\ \alpha = 0.05$.

For small values of the proportion $k/b$ smoothing by the mean value reduces the power of detecting a significant effect within such blocks because the effect is diluted. In this case, we suggest some solutions:

- One can replace the block mean by another summary statistic. We will see in the application some examples of appropriate statistics.
- One can choose a better segmentation.
- As noted before, even if the block size is very small ($b = 2$), the power of mean-BWA is larger than the power of the other methods used for comparison in this paper. So one can thus reduce the size of the smoothing windows $b$ to locate significant effects with more precision (e.g. if $b = 2$, a block is detected as significant means either exactly one of the two small regions is significant or both).

When $k/b$ exceeds a certain value $\left(\frac{k}{b}\right)^*$, mean-BWA is more powerful than SRW.

# Application to structural connectivity matrices of the human brain

In this section we apply mean-BWA to compare normalized whole-brain structural connectivity matrices derived from diffusion MRI tractography.

### *Brain connectivity matrices*

The processing pipeline producing the connection matrices use in this application is basically divided into two pathways. See Hagmann et al. (2008) for more details. On one hand, the cortical surface is extracted from a high resolution T1-weighted Magnetic Resonance (MR) image and subdivided into 66 anatomical parcels by matching the most important sulci using atlas-based segmentation. Each anatomical parcel is then subdivided into small cortical regions of interest (ROIs) of equal area. On the other hand, a whole brain tractography is performed on diffusion MR images, which results in millions of virtual fibers spread over the brain. The combination of these two procedures allows the construction of connectivity matrices by computing the connection density for each pairs of ROIs at each scale. The original cortical partition is made up of $N = 1000$ ROIs. However, the most suitable resolution between 1000 and 66 ROIs actually depends on the application. Thus, a hierarchical decomposition between 66 and 1000 ROIs is created by successive grouping. On the template brain, 2 or 3 neighboring ROIs at the 1000 scale are manually grouped into one ROI to build a partition into 483 ROIs. This grouping operation is repeated several times until the 66 parcels are recovered. This heuristic ends up with 5 embedded cortical parcellations with $N = \{1000, 483, 241, 133, 66\}$. Considering the cortical parcellation and the white matter tractography described above, the fiber bundle $F(k, l)$ connecting the pair of ROIs $(k, l)$ could be identified. The value of the connection matrix cell $M(k, l)$ is the connection density between these ROIs and is defined as follows:

$$M(k,l) = \left(\frac{2}{S(k) + S(l)}\right) \sum_{f \in F(k,l)} \frac{1}{l(f)},$$

where $S()$ is the surface of the ROI (●) and $l(f)$ is the length of the fiber $f$ along its trajectory. The correction term $l(f)$ in the denominator is needed to eliminate the linear bias towards longer fibers introduced by the tractography algorithm. We obtain at the end of the application of the pipeline a $N \times N$ symmetric matrix $M$. The pipeline is summarized in Figure 3.

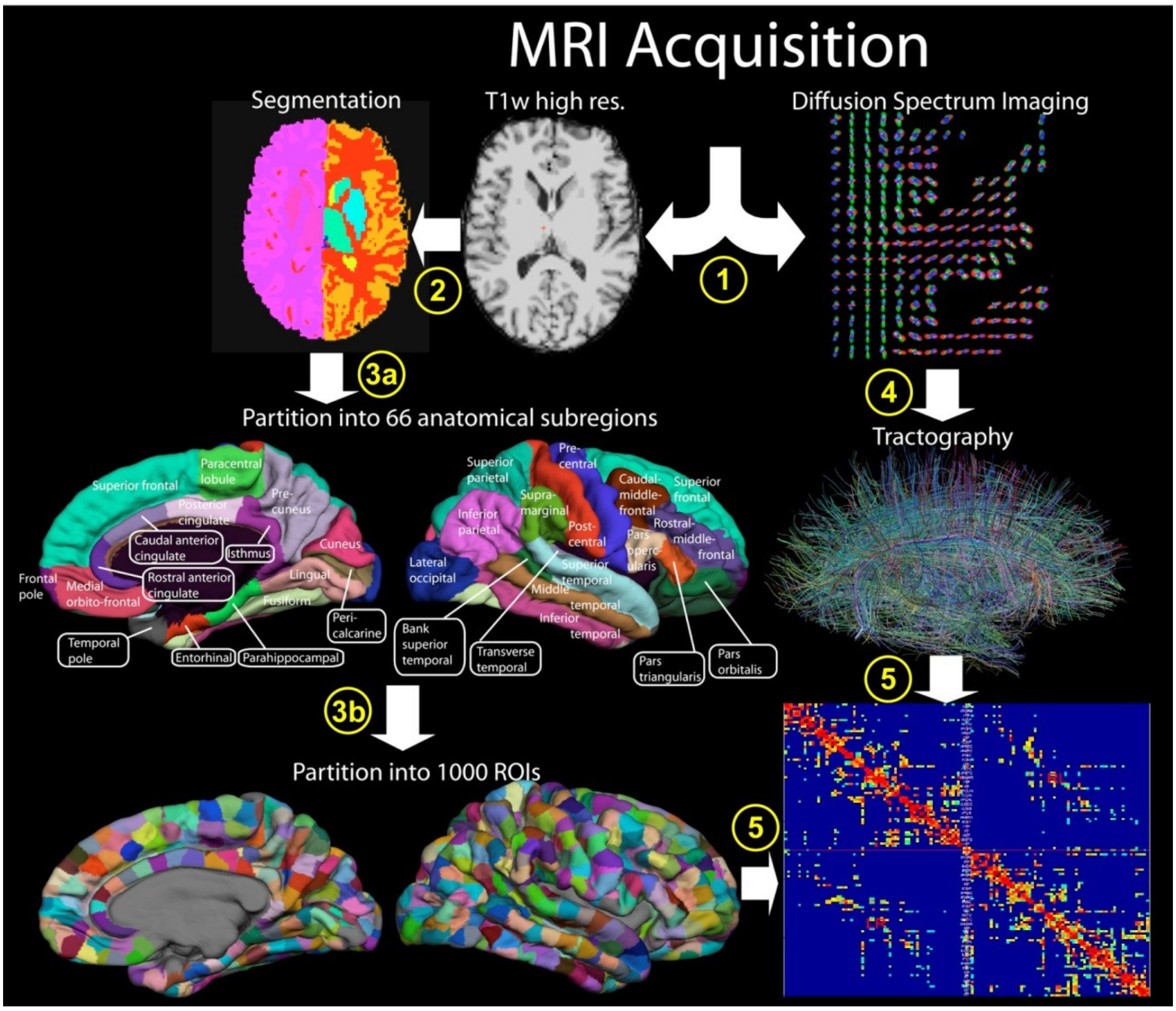

Figure 3: Extraction of a Whole Brain Structural Connectivity Matrix.

### *Description of the data*

The data we use in this application consists of $n_c = 15$ control connectivity matrices of resolution $N \times N = 483 \times 483$ which defines the small regions. From these 15 control matrices we generate $n_t = 15$ treatment matrices as follows:

For each small region $(k, l)$ we compute the mean $\bar{M}(k, l)$ and the standard deviation $s_{k,l}$ of the fiber density between the ROIs $k$ and $l$ over the 15 control connectivity matrices. We generate then 15 new values of $M(k, l)$ as random realizations of a normal distribution $N(\bar{M}(k, l), s_{k,l})$. We obtain 15 new connectivity matrices of size $N \times N = 483 \times 483$. Since the density should be a positive quantity, all negative values generated in this manner are set equal 0.

We randomly select a set of $m_1$ (the set of affected blocks) couples of ROIs from the resolution $66 \times 66$ which defines blocks (a block is a part of the connectivity matrix that represents the interconnections between two of the 66 anatomical brain regions). Since the connectivity matrix is symmetric, the total number of blocks is $\binom{66}{2}$. We add to each small region $(k, l)$ within these affected blocks of the 15 new matrices obtained in step 1 a random realization of a normal distribution $N(\Delta, s_{k,l})$ such that within each affected block $B_i$, only a proportion $k_i/b_i$ of small regions are affected. The number $\Delta$ is the raw effect. The block sizes $b_i$ of affected blocks as the proportions of affected small regions $k_i/b_i$ within these affected blocks are represented in Figure 4 as histograms.

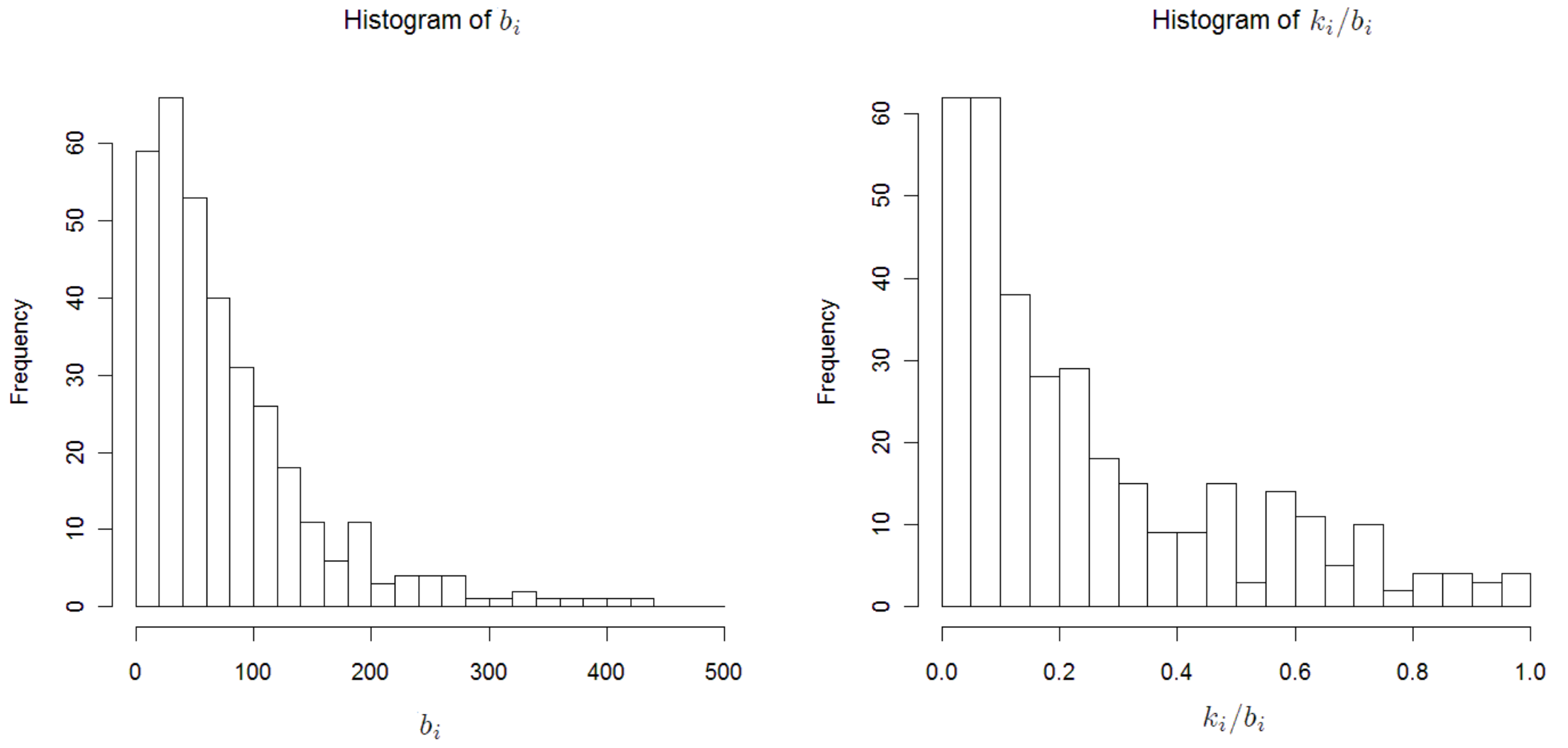

Figure 4: The distributions of the block sizes $b_i$ and the proportions of affected small regions within affected blocks $k_i/b_i$.

## *Data analysis and results*

We consider four strategies:

1. the small region-wise comparisons SRW described in the paper.
2. the mean-BWA where the summary statistic $t_i^1$ is the mean of the small regions in the block $B_i$.
3. the truncated mean-BWA, where the summary statistic in each block is defined by

$$t_i^2 = \left(\frac{1}{b_i}\right) \sum_{\{j \in I_{B_i}\}} I\{x_j > 0\},$$

where $I$ is the indicatrice function. This variable is interpreted as the number of interconnections between small regions within the block $B_i$. One can choose other thresholds of binarization depending on the nature of the problem. We use here 0 as a threshold for simplification. Since the distribution of the statistic $t_i^2$ is unknown, we use the Wilcoxon-Mann-Whitney rank test.

4. a bivariate statistic that includes both the mean and the truncated mean (Bivariate-BWA). For each block $B_i$, we define a bivariate statistic $\boldsymbol{t}$ as

$$\boldsymbol{t}_i = \begin{pmatrix} t_i^1 \\ t_i^2 \end{pmatrix} = \begin{pmatrix} \left(\frac{1}{b_i}\right) \sum_{\{j \in I_{B_i}\}} x_j \\ \left(\frac{1}{b_i}\right) \sum_{\{j \in I_{B_i}\}} I\{x_j > 0\} \end{pmatrix}.$$

The test follows by calculating the statistic $f$ as follows:

$$f = \left(\frac{n_c n_t}{n_c + n_t}\right) \left(t_i^1 - t_i^2\right)^T S^{-1} \left(t_i^1 - t_i^2\right) \left(\frac{n_c + n_t - 3}{2(n_c + n_t - 1)}\right),$$

where $S$ is the estimated covariance matrix of the data given by

$$S = \left(\frac{(n_c - 1)S_c + (n_t - 1)S_t}{(n_c + n_t - 2)}\right),$$

where $S_c$ and $S_t$ are the estimated covariance matrices of the control group and the treatment group respectively. The statistic $f$ follows a Fisher distribution $F_{\{2,(n_c+n_t-1)-2\}}$.

We apply the four strategies with different values of the raw effect $\Delta$. For correcting multiplicity we used the Bonferroni correction and the BH95 procedure. Because of the limitation of the BH95 procedure to independent or positively dependent tests we use also the Benjamini & Yekutieli procedure to control the FDR (BY01) for a general correlation structure that even tolerates negative correlations. See Dudoit and van der Laan (2008), Horn and Dunnett (2004). The level $\alpha$ is set to $0.05$ for all tests.

### Results and discussion

We summarize the percentage of significant small regions and blocks in Table 4.

| | | *SRW* | *mean-BWA* | *truncated mean-BWA* | *Bivariate-BWA* |
|---|---|---|---|---|---|
| $\Delta = 0.3$ | Bonferroni | 0 | 0.002 | 0.133 | 0.346 |
| | BH95 | 0 | 0.053 | 0.661 | 0.879 |
| | BY01 | 0 | 0 | 0.408 | 0.65 |
| $\Delta = 0.5$ | Bonferroni | 0 | 0.004 | 0.233 | 0.493 |
| | BH95 | 0.002 | 0.114 | 0.774 | 0.925 |
| | BY01 | 0 | 0.004 | 0.573 | 0.771 |
| $\Delta = 0.8$ | Bonferroni | 0 | 0.028 | 0.469 | 0.745 |
| | BH95 | 0.073 | 0.607 | 0.86 | 0.977 |
| | BY01 | 0 | 0.069 | 0.782 | 0.917 |
| $\Delta = 1$ | Bonferroni | 0 | 0.06 | 0.61 | 0.834 |
| | BH95 | 0.249 | 0.575 | 0.909 | 0.988 |
| | BY01 | 0.003 | 0.225 | 0.863 | 0.954 |
| $\Delta = 1.5$ | Bonferroni | 0 | 0.349 | 0.883 | 0.961 |
| | BH95 | 0.909 | 0.987 | 0.957 | 0.998 |
| | BY01 | 0.385 | 0.848 | 0.951 | 0.992 |

Table 4: The average power of the competing methods for different values of the raw effect $\Delta$.

We see clearly that SRW is ineffective when the raw effect is small. As expected, both FDR procedures give better results than the FWER procedures. However, the BH95 has higher power than BY01, but its application depends on the joint distribution of the tests. For more details see Horn and Dunnett (2004), Dudoit and van der Laan (2008). On the other hand, the block strategy used with different summary statistics (mean-BWA, truncated mean-BWA, Bivariate-BWA) shows a real advantage, in particular for the multivariate summary statistic. We expect better results if other appropriate summary statistics that include additional information like fractional anisotropy, diffusion tensor orientation, etc., are used. We expect also that the use of network measures in this context such as centrality, modularity, etc. as summary statistics would give better results. See Bullmore and Sporns (2009) for a list of relevant network measures that could be used in this application.

We can summarize the procedure we propose in this paper as follows.

1. Choose a segmentation that defines blocks.
2. Mask all small regions which are not considered in comparison.
3. Choose an appropriate summary statistic to be used in multiple comparisons (univariate or multivariate).
4. Choose a multiple comparisons procedure (that controls FWER or FDR).
5. Make a local investigation in detected significant blocks if desired and if this makes sense.

# Conclusion

We have proposed a strategy that improves the power of a family of tests based on reducing the number of tests by grouping tests into blocks. The most important aspect guaranteeing the success of this strategy is that the blocks are defined by the nature of the problem at hand and that the small regions within blocks are likely to be affected simultaneously. We have shown in the simulation examples and in the application the relevance in neuroimaging.

It should be emphasized that the block strategy as presented in this paper uses some particular examples of summary statistics and has shown a real advantage over the small region-wise comparisons even if the blocks are not completely affected. However, depending on the nature of the problem at hand, the choice of the summary statistic is of crucial importance. In our application, a connectivity matrix defines a network (Hagmann et al., 2008) and hence a block of the connectivity matrix represents a sub-network. In this case, one could use network attributes such as node-degree distribution, centrality, efficiency, small worldness, etc. of each sub-network as a summary statistic of the correspondent block. Better results could be obtained also by using improved methods like the use of the empirical null distribution. See Schwartzman et al. (2009).

### *Acknowledgement*

This work was supported by the Swiss National Science Foundation under grant number 205321_121945 and by the Centre d'imagerie BioMédicale (CIBM) of the University of Lausanne (UNIL), the Swiss Federal Institute of Technology Lausanne (EPFL), the University of Geneva (UniGe), the Centre Hospitalier Universitaire Vaudois (CHUV), the Hôpitaux Universitaire de Genève (HUG) and the Leenaards and the Jeantet Foundations.